\documentclass[12pt]{article}
\usepackage{amssymb, amsmath}

\def \AA {\mathcal{A}}
\def \BB {\mathcal{B}}
\def \FF {\mathcal{F}}
\def \HH {\mathcal{H}}
\def \SS {\mathcal{S}}

\def \CCC {\mathbb{C}}
\def \NNN {\mathbb{N}}
\def \RRR {\mathbb{R}}

\def \tr {\mathrm{Tr}}

\def \bm {\mathbf{m}}
\def \bn {\mathbf{n}}

\def \bv {\mathrm{v}}

\def \equals {\ = \ }

\newtheorem{proposition}{Proposition}

\title{Examples of bosonic de Finetti states over finite dimensional Hilbert spaces}
\author{Alex D. Gottlieb}
\date{}
\begin{document}

\maketitle
\begin{abstract}
According to the Quantum de Finetti Theorem, locally normal
infinite particle states with Bose-Einstein symmetry can be
represented as mixtures of infinite tensor powers of vector
states.  This note presents examples of infinite-particle states
with Bose-Einstein symmetry that arise as limits of Gibbs
ensembles on finite dimensional spaces, and displays their de
Finetti representations.  We consider Gibbs ensembles for systems
of bosons in a finite dimensional setting and discover limits as
the number of particles tends to infinity, provided the
temperature is scaled in proportion to particle number.
\end{abstract}

\section{Introduction}
According to the Quantum de Finetti Theorem \cite{HudsonMoody},
locally normal infinite-particle states with Bose-Einstein
symmetry can be represented as mixtures of infinite tensor powers
of vector states. This note presents examples of infinite-particle
states with Bose-Einstein symmetry that arise as limits of Gibbs
ensembles on finite dimensional spaces, and displays their de
Finetti representations.

The central example is as follows.  If the single-particle Hilbert
space $\HH$ is finite dimensional, the projector onto the
symmetric subspace of the $n$-particle space can be normalized,
and this defines the infinite-temperature ensemble for $n$ bosons
with single-particle space $\HH$.  For each fixed $m \in \NNN$,
the $m$-particle reduced density operators under the $n$-boson
infinite-temperature ensembles converge, as $n$ tends to infinity,
to the density operator describing the $m$-particle statistics
under a certain bosonic infinite-particle state $\omega_0$.  The
infinite-particle state $\omega_0$ has a de Finetti representation
as a mixture of infinite tensor powers of vector states
$P_{\mathrm{v}}$, where $\mathrm{v}$ is a unit vector and
$P_{\mathrm{v}}=|\mathrm{v} \rangle \langle \mathrm{v}|$ denotes
the projector onto the span of $\mathrm{v}$. In the de Finetti
mixture for $\omega_0$, the weight of the tensor power state
${P_{\mathrm{v}}}^{\otimes \infty}$ is the probability density for
$\mathrm{v} \in \HH \stackrel{\sim}{=} \CCC^{d+1}$ to equal
\begin{equation}
\label{v}
   \mathrm{v}(p,\theta) \equals \big( e^{i\theta_0}\sqrt{p_0},\ e^{i\theta_1}\sqrt{p_1},\ldots,\ e^{i\theta_d}\sqrt{p_d}\big)
\end{equation}
when $p=(p_0,p_1,\ldots,p_d)$ is sampled uniformly from the
$d$-dimensional simplex $\Delta_d$ and the phase angles $\theta_i$
in $\theta=(\theta_0,\theta_1,\ldots,\theta_d)$ are each sampled
uniformly from $[0,2\pi)$, independently of one another and of
$p$. Thus the infinite-particle state $\omega_0$ corresponds to
the uniform probability measure on $\Delta_d\times
[0,2\pi)^{d+1}$.

Similar limits are obtained for finite temperature Gibbs
ensembles, provided the temperature is scaled properly. Suppose
$H$ is a Hermitian operator on the single-particle space $\HH
\stackrel{\sim}{=} \CCC^{d+1}$ and $\Gamma_n(\beta)$ denotes the
Gibbs canonical ensemble for $n$ noninteracting bosons with
single-particle Hamiltonian $H$ at inverse temperature $\beta$.
Then, as $n$ tends to infinity, the $m$-particle reduced density
operators under $\Gamma_n(\beta/n)$ converge to the $m$-particle
density of a certain bosonic infinite-particle state
$\omega_{\beta}$. The infinite-particle state $\omega_{\beta}$ is
an average of states ${P_{\mathrm{v}}}^{\otimes \infty}$ with
respect to the probability density on $\Delta_d\times
[0,2\pi)^{d+1}$ that minimizes the ``free energy"
\[
   \int\limits_{ [0,2\pi)^{d+1}} \int_{\Delta_d}
    \langle \mathrm{v}, H \mathrm{v} \rangle f(p,\theta) \hspace{2pt} dp \hspace{2pt} d\theta
    \ + \ \frac{1}{\beta} \int\limits_{ [0,2\pi)^{d+1}} \int_{\Delta_d}
    f(p,\theta) \ln f(p,\theta) \hspace{2pt} dp \hspace{2pt} d\theta \ ,
\]
where $\mathrm{v}=\mathrm{v}(p,\theta)$ is as in (\ref{v}). We
obtain similar results for bosons with ``mean field" interactions,
but again we must scale temperature in proportion to the number of
particles.  This stands in contrast to the analogous mean field
limits for {\it distinguishable} particles, which are obtained
without any peculiar scaling of temperature
\cite{FannesSpohnVerbeure}.

The physical relevance of these facts is limited.  On the one
hand, they concern limits of canonical ensembles, which are
appropriate when the number of bosons is fixed, and therefore not
appropriate for massless bosons (e.g., photons).  On the other
hand, massive bosons inhabit infinite dimensional Hilbert spaces,
so to speak, whereas our results concern finite dimensional
Hilbert spaces. However, the sort of ensemble we study is 
appropriate for (noninteracting) systems of $n$ material bosons in
thermal equilibrium, {\it in case it is known that every one of
these bosons is trapped in a potential well of depth $E$}. The
statistical state of that system would be a {\it conditional}
Gibbs ensemble, supported on the finite dimensional Hilbert space
spanned by the symmetrized products of trapped (bound) states.
Only {\it noninteracting} systems of trapped bosons are
considered, because the conditional Gibbs ensemble only makes
sense if the Hamiltonian of the system commutes with the
observable that every particle is trapped.

Our results are presented in Section~\ref{results}, after a quick
review of the Quantum de Finetti Theorem in the next section.

\section{The Quantum de Finetti Theorem}
\label{quantumDeFinetti}

Let $\HH$ be Hilbert space (which we will call the single-particle
Hilbert space) and let $\HH^{\otimes n}$ denote the $n$-fold
tensor power of $\HH$ (the $n$-particle Hilbert space). When $\pi$
denotes a permutation of $\{1,2,\ldots,n\}$, let $U_{\pi}$ denote
the unitary ``permutation" operator on $\HH^{\otimes n}$ defined
by
\[
    U_{\pi}(x_1 \otimes x_2 \otimes \cdots \otimes x_n) \equals x_{\pi(1)}
    \otimes x_{\pi(2)} \otimes \cdots \otimes x_{\pi(n)}\ .
\]
For each $n \in \NNN$ let $D_n$ be a density operator on the
$n$-particle Hilbert space $\HH^{\otimes n}$, the $n$-fold tensor
power of $\HH$. We want the density operators $D_n$ to be
symmetric, and we assume
\begin{trivlist}
\item{(A)}  for all $n$, the density operator $D_n$ commutes with
any permutation operator $U_{\pi}$ on $\HH^{\otimes n}$ \ .
\end{trivlist}
We are especially interested here in systems of bosons, for which
\begin{trivlist}
\item{(B)}  for all $n$,  $D_n U_{\pi} = D_n$ for any permutation
operator $U_{\pi}$ on $\HH^{\otimes n}$.
\end{trivlist}
Condition (B) is stronger than (A). We also want the sequence
$\{D_n\}$ of density operators to be consistent with respect to
``subsampling" in the sense that
\begin{trivlist}
\item{(C)}  for all $m < n$, $D_{n:m}=D_m$,
\end{trivlist}
where $D_{n:m}$ denotes the $m^{th}$ partial trace of $D_n$, i.e.,
the operator such that
\[
    \tr(D_{n:m} A ) \equals \tr( D_n (A\otimes I \stackrel{n-m\ times}{\otimes \cdots \otimes} I))
\]
for all $A \in \BB(\HH^{\otimes m})$.

The structure of sequences $\{D_n\}$ of density operators
satisfying (C) and (A) or (B) is given by the quantum analogue of
the de Finetti Theorem of probability theory \cite{HudsonMoody}.
Let $\rho$ be a density operator on $\HH$. A sequence $\{D_n\}$ of
density operators of the form
\begin{eqnarray}
\label{product}
     D_1 & = &  \rho \nonumber \\
     D_2 & = & \rho \otimes \rho \nonumber \\
     D_3 & = & \rho \otimes \rho \otimes \rho \ ,\quad \hbox{et cetera}
\end{eqnarray}
always satisfies (A) and (C), but it satisfies (B) and (C) if and
only if $\rho$ is a pure state, i.e., a rank one projector on
$\HH$. Roughly speaking, any sequence of density operators
satisfying (A) and (C) is uniquely representable as a mixture of
sequences of the form (\ref{product}).  That is, if $\{D_n\}$
satisfies (A) and (C) then there exists a unique probability
measure $\mu$ supported on the single-particle density operators
such that
\begin{equation}
\label{integral0}
      D_n \equals\int \rho^{\otimes n} \mu(d\rho)
\end{equation}
for all $n$.  Furthermore, if $\{D_n\}$ satisfies (B) and (C),
then the measure $\mu(d\rho)$ in the integral representation
(\ref{integral0}) is even supported on the set of vector states
$\rho=P_{\psi}$. This paraphrases the propositions of
\cite{HudsonMoody}, ignoring the technical details; we now restate
the results with more care.

For $m \le n$, let $j_{mn}$ denote the *-isomorphic embedding
\[
    j_{mn}(B) \equals B \otimes I^{\otimes n-m}
\]
of $\BB(\HH^{\otimes m})$ into $\BB(\HH^{\otimes n})$.  The system
of C* algebras $\BB(\HH^{\otimes n})$ and isomorphic injections
$j_{mn}$ has an inductive limit $\AA$. The inductive or direct
limit in the category of C* algebras may be constructed as in
\cite[Proposition 11.4.1]{KadisonRingrose}.  The inductive limit
$\AA$ is unique up to isomorphism, and for each $n$ there is a
*-isomorphism $i_n$ from $\BB(\HH^{\otimes n})$ into $\AA$ such
that $i_n j_{mn}=i_m$ for all $m \le n$ and the union of the
images $i_n(\BB(\HH^{\otimes n}))$ is dense in $\AA$. A sequence
$\{D_n\}$ of density operators satisfying the conditions (C) can
be used to define a continuous positive linear functional $\omega$
on $\AA$ by
\begin{equation}
\label{omega}
     \omega( i_n (B) ) \equals\tr( D_n B) \qquad \qquad \forall  B \in \BB(\HH^{\otimes n}) \ .
\end{equation}
This is well-defined thanks to the consistency conditions (C) and
the density of $\cup i_n(\BB(\HH^{\otimes n}))$ in $\AA$. In
particular, $\omega(e)=1$, where $e$ is the identity element of
the C* algebra $\AA$.  If $\{D_n\}$ satisfies (A) as well as (C)
then $\omega$ is symmetric in the sense that
\begin{equation}
\label{symmetric}
   \omega (i_n (U_{\pi} B U^*_{\pi}) ) \equals\omega ( i_n ( B  ) )
\end{equation}
for all $n$, all $B \in \BB(\HH^{\otimes n})$, and all $\pi \in
\Pi_n$, the set of permutations of $\{1,2,\ldots,n\}$.   The set
of all ``symmetric states" on $\AA$, i.e., the set
\[
     \SS \equals\Big\{ \omega \in \AA^* \ \Big| \ \omega(e)=1
     \ \hbox{and} \ \omega(x^*x) \ge 0 \ \forall x \in \AA
     \ \hbox{and} \ \omega \ \hbox{satisfies} \ (\ref{symmetric}) \Big\}\ ,
\]
is a convex subset of the Banach dual $\AA^*$ of $\AA$, and it is
compact with respect to the weak* topology. Let $\SS_1$ denote the
space of single-particle states, i.e., the set
\[
     \SS_1 \equals\Big\{ \rho \in \BB(\HH)^* \ \Big| \ \rho(I)=1
     \ \hbox{and} \ \omega(A^*A) \ge 0 \ \forall A \in \BB(\HH) \Big\}
\]
endowed with the relative weak* topology it inherits as a subset
of the Banach dual $\BB(\HH)^*$ of $\BB(\HH)$ . It was first shown
in \cite{Stormer} that each $\omega \in \SS$ has a unique
representation as an integral of product states
\begin{equation}
\label{integral}
   \omega \equals \int_{\SS_1}  \rho \otimes \rho \otimes \rho \otimes \cdots \cdots \ \mu(d\rho) \equals
   \int_{\SS_1}  \rho^{\otimes \infty}  \mu(d\rho)\ ,
\end{equation}
where $\mu$ is a probability measure on the $\sigma$-algebra
$\FF_1$ generated by the intersections with $\SS_1$ of weak* open
sets in $\BB(\HH)^*$. We sketch a proof of this, following
reference \cite{HudsonMoody}: First, the extreme points of $\SS$
are identified as the product states $\rho^{\otimes\infty} $.
Thus, the set of extreme points is the image of the compact space
$\SS_1$ under the continuous injection $\rho \longmapsto \rho^{
\otimes\infty} $, and it follows that the extreme set is closed in
$\SS$. The existence of an integral representation
(\ref{integral}) is then a consequence of the Krein-Milman
Theorem, and its uniqueness is shown in \cite{HudsonMoody} by a
direct argument.

It is further shown in \cite{HudsonMoody} that the measure
$\mu(d\rho)$ appearing in the integral representation
(\ref{integral}) of $\omega$ is supported on the measurable subset
of {\it normal} states on $\BB(\HH)$ if $\omega$ is determined, as
in formula (\ref{omega}) above, by sequences of density operators
satisfying (A) and (C).  If, in addition, the sequence of density
operators defining $\omega$ satisfies (B), then the measure
$\mu(d\rho)$ is even supported on the vector states
$\rho(A)=\langle \psi, A \psi \rangle$ with $\|\psi\|=1$.

\section{Examples of bosonic de Finetti states}
\label{results}

In this section we exhibit some sequences $\{D_n\}$ satisfying (B)
and (C) that are obtained from natural statistical ensembles.   In
all of these examples, the single-particle Hilbert space $\HH$ is
finite dimensional.  After introducing the notation, we will state
all of our results before proceeding to their proofs.

Let $\HH=\CCC^{d+1}$ and let $\HH^{(n)}$ denote the subspace of
symmetric vectors in $\HH^{\otimes n}$. Let $\Sigma_n$ denote the
symmetrizing projector
\begin{equation}
\label{Sigma-n}
   \Sigma_n \equals\frac{1}{n!} \sum_{\pi \in \Pi_n} U_{\pi}
\end{equation}
from $\HH^{\otimes n}$ onto $\HH^{(n)}$.   We now introduce
notation for the occupation number basis of $\HH^{(n)}$ relative
to a fixed orthonormal (ordered) basis $\{e_j\}$ of $\HH$. Let
$\bn = (n_0,n_1,\ldots,n_d)$ be an ordered $d+1$-tuple of
nonnegative integers (occupation numbers) and let $\#\bn$ denote
$\sum n_j$.    We use the notation
\[
    \binom{n}{\bn} \equals n! \Big/ \prod\limits_{i=0}^d n_i!
\]
for multinomial coefficients. The vector
\begin{equation}
\label{Psi_n} \Psi_{\bn} \equals\sqrt{\binom{n}{\bn}}\
\Sigma_n(e_0^{\otimes n_0} \otimes e_1^{\otimes n_1} \otimes
\cdots \otimes e_d^{\otimes n_d} )
\end{equation}
is a unit vector in $\HH^{(n)}$, and the set of vectors
$\{\Psi_{\bn} \ | \ \#\bn = n\}$ is an orthonormal basis of
$\HH^{(n)}$. Let $P_{\bn}$ denote the rank-one projector onto the
span of $\Psi_{\bn}$:
\begin{equation}
       P_{\bn}\Phi \equals\langle \Psi_{\bn}, \Phi\rangle \Psi_{\bn}\ .
\label{Pn}
\end{equation}

We begin by considering the ``uniformly mixed" density operators
supported on $\HH^{(n)}$:
\begin{proposition}
\label{uniform1} Let $\Sigma_n$ denote the symmetrizing projector
(\ref{Sigma-n}). For each $m$,
\begin{equation}
\label{limitUniform}
       S_m \ \equiv \ \lim\limits_{n\rightarrow \infty} \frac{1}{\tr\Sigma_n}\Sigma_{n:m} \equals
    \sum_{\bm: \#\bm=m} \Big\{
    \binom{m}{\bm} \int_{\Delta_d} \prod_{i=0}^d p_i^{m_i} \lambda_d(dp)
    \Big\}
    P_{\bm}\ ,
\end{equation}
where $\lambda_d(dp)$ denotes normalized Lebesgue measure on the
$d$-dimensional simplex
\[
\Delta_d \equals \Big\{   p=(p_0,p_1,\ldots,p_d) \in \RRR^{d+1} \
\big| \  0 \le p_i \quad i = 1,2,\ldots,d \quad \hbox{and} \quad
\sum_{i=0}^d p_i =1 \Big\}\ .
\]
\end{proposition}
The sequence $\{S_m\}$ satisfies (B) and (C) of
Section~\ref{quantumDeFinetti}. By the Quantum de Finetti Theorem,
there exists a measure $\mu$ supported on the pure states on
$\CCC^{d+1}$ such that
\[
    S_m \equals\int P^{\otimes m} \mu(dP)
\]
for all $m \in \NNN$.   This measure can be described as follows.
 Define the map
\[
    \bv:\Delta_d\times [0,2\pi)^{d+1} \longrightarrow \CCC^{d+1}
\]
 by
\begin{equation}
\label{bv}
    \bv(p_0,p_1,\ldots,p_d,\theta_0,\theta_1,\ldots,\theta_d) \equals\sum_{j=0}^d e^{i\theta_j}\sqrt{p_j} \ e_j
\end{equation}
where $\{e_i\}$ is the standard basis of $\CCC^{d+1}$. The map
$\bv$ is many-one onto the set of unit vectors in $\CCC^{d+1}$.
The probability measure $\mu(dP)$ is the one induced via $\bv$
from the uniform measure
\[
    \lambda(dp)\sigma(d\theta)
    \ \equiv \ \lambda(dp)\ \frac{d\theta_0}{2\pi}\frac{d\theta_1}{2\pi}\cdots\frac{d\theta_d}{2\pi}
\]
on $\Delta_d \times [0,2\pi)^{d+1}$.  In other words,
\begin{proposition}
\label{uniform2} The density operator (\ref{limitUniform}) equals
\begin{equation}
\label{integralRepresentationUniform}
       \int_{\Delta_d}  \int_{[0,2\pi)^{d+1}}
               \big(
           P_{\bv(p,\theta)}
                 \stackrel{
              \hbox{m times}
              }{
               \otimes   \cdots \otimes
               }
                 P_{\bv(p,\theta)}  \big) \sigma(d\theta)  \lambda_d(dp) \ .
\end{equation}
\end{proposition}

Next we consider Gibbs ensembles for noninteracting systems of
bosons.  Let
\begin{equation}
    H_n \equals \sum_{i = 1}^n T_i
\label{noninteractingHam}
\end{equation}
be the Hamiltonian for $n$ noninteracting bosons with
single-particle space $\HH=\CCC^{d+1}$. Let $\{e_j\}$ be an
orthonormal basis of $\HH$ consisting of eigenvectors of the
single-particle operator $T$, so that $Te_j=\epsilon_j e_j$. The
Gibbs density operator for the $n$ boson system is
\begin{equation}
    \Gamma_n(\beta) \equals \frac{1}{Z_{n,\beta}} \sum_{\bn: \#\bn=n}  \prod_{i=0}^d e^{-\beta n_i \epsilon_i} P_{\bn}
\qquad \mathrm{with} \qquad
    Z_{n,\beta} \equals\sum_{\bn: \#\bn=n}  \prod_{i=0}^d e^{-\beta n_i \epsilon_i}\ .
\label{noninteractingGibbs}
\end{equation}
An interesting limit is attained if temperature is scaled in
proportion to $n$ as $n \longrightarrow \infty$. If the
temperature is {\it not} scaled as $n \longrightarrow \infty$ then
a sort of Bose-Einstein condensation is attained in the limit.
\begin{proposition}
\label{noninteracting} Let $H_n$ be the noninteracting Hamiltonian
(\ref{noninteractingHam}) and let $\Gamma_n(\beta)$ denote the
Gibbs density (\ref{noninteractingGibbs}).  Let $\{e_j\}$ be an
orthonormal basis of $\HH$ consisting of eigenvectors of the
single-particle operator $T$, so that $Te_j=\epsilon_j e_j$.
\begin{trivlist}
\item{(i)}\quad For each $m \in \NNN$, the limit $\lim\limits_{n
\rightarrow \infty} \Gamma_{n:m}(\beta/n)$ exists and equals
\[
    \sum_{\bm: \#\bm=m}  \Big\{
    \binom{m}{\bm}
    \int_{\Delta_d} \prod_{i=0}^d p_i^{m_i}
    Z_{\beta}^{-1}\prod_{i=0}^d e^{-\beta \epsilon_i p_i} \lambda_d(dp)
    \Big\} P_{\bm}
\]
with $Z_{\beta}^{-1}=\int_{\Delta_d} \prod_{i=0}^d \exp(-\beta
\epsilon_i p_i) \lambda_d(dp)$. \item{(ii)}\quad If $\epsilon_0 <
\epsilon_1 \le \cdots \le \epsilon_d$, then for each $m \in \NNN$,
\[
    \lim_{n \rightarrow \infty} \Gamma_{n:m}(\beta) \equals P_{(m,0,\ldots,0)} \equals {P_{e_0}}^{\otimes m}\ .
\]
\end{trivlist}
\end{proposition}

Finally, we consider systems with two-particle interactions in the
``mean field" scaling. Let $V$ be a Hamiltonian operator on
$\HH\otimes \HH$ such that $V(x\otimes y)=V(y\otimes x)$ for all
$x,y\in \HH$. For $n>2$, define the Hamiltonian
\begin{equation}
    H_n \equals \sum_{i =1}^n T_i \ + \ \frac{1}{n-1}\sum_{1 \le i < j \le n} V_{ij}\ ,
\label{meanFieldHam}
\end{equation}
where $V_{ij}$ denotes the operator obtained by applying $V$ to
the $i^{th}$ and $j^{th}$ factors of $\HH^{\otimes n}$. For any $n
\in \NNN$ and any $\beta \in \RRR$, the $n$-particle Gibbs density
at inverse temperature $\beta$ for the Hamiltonian
(\ref{meanFieldHam}) is
\begin{equation}
       \Gamma_n(\beta) \equals\frac{1}{\tr( e^{- \beta H_n} \Sigma_n)} \ e^{- \beta H_n} \Sigma_n \ .
\label{Gibbs}
\end{equation}

\begin{proposition}
\label{canonical} Let $\Gamma_n(\beta)$ denote the Gibbs density
(\ref{Gibbs}).  For each $m$, the limit
\[
        G_m \equals \lim_{n\rightarrow \infty} \big\{ \Gamma_n(\beta/n) \big\}_{:m}
\]
exists and defines a density operator on $(\CCC^{d+1})^{\otimes
n}$. The de Finetti representation of $G_m$ is
\[
    \frac{1}{Z_{\beta}}
       \int_{\Delta_d}  \int_{[0,2\pi)^{d+1}}
             \stackrel{ \hbox{m times} }{ P_{\bv} \otimes \cdots \otimes P_{\bv}}
          e^{ -\beta \{ \tr(T P_{\bv}) +  \tr(V (P_{\bv} \otimes P_{\bv}))/2 \} }
      \sigma(d\theta) \lambda_d(dp)
\]
with $\bv=\bv(p,\theta)$ as in (\ref{bv}) and
\[
Z_{\beta} =
       \int_{\Delta_d}   \int_{[0,2\pi)^{d+1}}
            e^{ -\beta \{ \tr(T P_{\bv}) +  \tr(V (P_{\bv} \otimes P_{\bv}))/2 \} }
       \sigma(d\theta)  \lambda_d(dp)  \ .
\]
\end{proposition}

The rest of this section is devoted to proving the above
propositions.

Recall the definition (\ref{Pn}) of the projectors $P_{\bn}$. For
each $n \in \NNN$, let $\rho_n$ be an $n$-particle density
\[
       \rho_n \equals\sum_{\bn: \#\bn=n} w_n(\bn) P_{\bn} \ ,
\]
where $w_n$ is a probability measure on the set $\{\bn|
\#\bn=n\}$.  Each probability measure $w_n$ can be associated with
the discrete probability measure
\[
     \omega_n \equals\sum_{\bn: \#\bn=n} w_n(\bn) \delta(p-\tfrac{1}{n}\bn)
\]
on the $d$-dimensional simplex $ \Delta_d $. It may be verified
that
\[
    P_{\bn:m} \equals \binom{n}{m}^{-1}\sum_{\bm: \#\bm=m}
    \prod\limits_{i=0}^d \binom{n_i}{m_i}     P_{\bm}
\]
(this equals $0$ if any $m_i > n_i$ for any $i$), and therefore
\begin{eqnarray}
       \rho_{n:m} & = &
       \binom{n}{m}^{-1}\sum_{\bm: \#\bm=m}
    \Big[  \sum_{\bn: \#\bn=n} w_n(\bn) \prod\limits_{i=0}^d \binom{n_i}{m_i} \Big]    P_{\bm}
    \nonumber \\
    & = &
      \sum_{\bm: \#\bm=m}  \binom{m}{\bm}
    \Bigg[  \sum_{\bn: \#\bn=n} w_n(\bn)
    \frac{
        \prod_{i=0}^d \frac{n_i}{n}(\frac{n_i}{n} - \frac1n ) \cdots (\frac{n_i}{n} - \frac{m_i-1}{n})
        }
        {
        1(1-\tfrac1n)(1-\tfrac2n) \cdots (1 - \tfrac{m-1}{n})
    }
    \Bigg]    P_{\bm}  \ . \nonumber \\
    \label{rewriteMe}
\end{eqnarray}
The coefficient of $P_{\bm}$ in (\ref{rewriteMe}) may be written
\[
        \binom{m}{\bm} \int_{\Delta_d} f_n(p) \omega_n(dp) \ ,
\]
where
\[
    f_n(p) \equals 1\!\mathrm{l}_{\{p_i>(m_i-1)/n \ \forall i\}}(p) \frac{
        \prod_{i=0}^d p_i(p_i - \frac1n ) \cdots (p_i - \frac{m_i-1}{n})
        }
        {
        1(1-\tfrac1n)(1-\tfrac2n) \cdots (1 - \tfrac{m-1}{n})
    }  \ .
\]
The functions $f_n(p)$ converge uniformly to $
    \prod_{i=0}^d p_i^{m_i}
$ on $\Delta_d$.  Therefore, if $\omega_n$ converges weakly to
some probability measure $\omega(dp)$ on $\Delta_d$, then
\begin{equation}
\label{limitFormula}
    \lim_{n \rightarrow \infty} \rho_{n:m} \equals\sum_{\bm: \#\bm=m}  \binom{m}{\bm} \int_{\Delta_d} \prod_{i=0}^d p_i^{m_i} \omega(dp) \ P_{\bm}\ .
\end{equation}

The probability measures on $\Delta_d$ corresponding to the Gibbs
density operators (\ref{noninteractingGibbs}) for noninteracting
bosons are
\begin{equation}
\label{noninteractingCanonicalOmegas}
    \omega_n \equals  Z_{n,\beta}^{-1} \sum_{\bn: \#\bn=n} \prod_{i=0}^d e^{-\beta n_i \epsilon_i}
    \delta(p-\tfrac{1}{n}\bn )\ .
\end{equation}
If all of the eigenvalues of $T$ are equal, then the measures
(\ref{noninteractingCanonicalOmegas}) converge weakly to
$\lambda_d(dp)$, the uniform probability measure on the simplex,
but if $\epsilon_0$ is strictly smaller than all of the other
eigenvalues of $T$, then the measures
(\ref{noninteractingCanonicalOmegas}) converge weakly to
$\delta(p-(1,0,\ldots,0) )$, a point-mass at the lowest energy
vertex of the simplex.    This convergence implies
Propositions~\ref{uniform1} and assertion (ii) of
Proposition~\ref{noninteracting} by formula (\ref{limitFormula}).
On the other hand, the probability measures corresponding to the
Gibbs density operators $\Gamma_n(\beta/n)$ for noninteracting
bosons are
\[
    \omega_n \equals  Z_{n,\beta}^{-1} \sum_{\bn: \#\bn=n} \prod_{i=0}^d e^{-\beta  \epsilon_i n_i/n}
    \delta(p-\tfrac{1}{n}\bn )\ ,
\]
and these converge weakly to
\[
 Z_{\beta}^{-1} \prod_{i=0}^d e^{-\beta \epsilon_i p_i} \lambda_d(dp)
\]
with $Z_{\beta}=\int_{\Delta_d} \prod_{i=0}^d\exp(-\beta
\epsilon_i p_i) \lambda_d(dp)$. This proves assertion (i) of
Proposition~\ref{noninteracting}.


To prove Proposition~\ref{uniform2}, we will show that
(\ref{integralRepresentationUniform}) and (\ref{limitUniform}) are
equal.  Define the rank-one operators $Q_{jk}(x)= \langle e_k, x
\rangle e_j$. From (\ref{bv}),
\[
    P_{\bv(p,\theta)} \equals\sum_{j,k=0}^d e^{i(\theta_j-\theta_k)}\sqrt{p_jp_k} \ Q_{jk}
\]
and therefore ${ P_{\bv(p,\theta)} }^{\otimes m}$ equals
\begin{equation}
\label{doubleSum}
     \sum_{j_1,\ldots,j_m=0}^d \sum_{k_1,\ldots,k_m=0}^d \prod\limits_{r=0}^d
     \sqrt{p_{j_r}p_{k_r}} e^{i(\theta_{j_r}-\theta_{k_r})}
     \ Q_{{j_1}{k_1}} \otimes Q_{{j_2}{k_2}} \otimes \cdots \otimes Q_{{j_m}{k_m}} \ .
\end{equation}
For $i=0,1,\ldots,d$, let $N_i:\{0,1,\ldots,d\}^m \longrightarrow
\NNN$ be defined by
\[
    N_i(x_1,x_2,\ldots,x_m) \equals \#\big\{k \in \{1,2,\ldots,m\}:\ x_k = i \big\}
\]
and define
\[
    N(x_1,x_2,\ldots,x_m) \equals \big(N_0(x_1,x_2,\ldots,x_m),\ldots,N_d(x_1,x_2,\ldots,x_m)\big)\ .
\]
If $N(j_1,\ldots,j_m)=N(k_1,\ldots,k_m)$ then
\[
    \int_{[0,2\pi)^{d+1}}\prod\limits_{r=0}^d e^{i(\theta_{j_r}-\theta_{k_r})}\sigma(d\theta) \equals 1\ ,
\]
but otherwise it equals $0$.  Thus, from (\ref{doubleSum}),
\begin{eqnarray*}
    & &
     \int_{\Delta_d} \int_{[0,2\pi)^{d+1}}
     { P_{\bv(p,\theta)} }^{\otimes m}
     \sigma(d\theta)\lambda_d(dp) \\
     & = &
     \sum_{\bm:\#\bm = m} \int_{\Delta_d}\prod\limits_{i=0}^m p_i^{m_i} \lambda_d(dp)
    \sum_{\stackrel{j_1,\ldots,j_m:}{N(j_1,\ldots,j_m)=\bm}}
    \sum_{\stackrel{k_1,\ldots,k_m:}{N(k_1,\ldots,k_m)=\bm}}
     Q_{{j_1}{k_1}}  \otimes \cdots \otimes Q_{{j_m}{k_m}} \\
     & = &
     \sum_{\bm:\#\bm = m} \int_{\Delta_d}\prod\limits_{i=0}^m p_i^{m_i} \lambda_d(dp)
     \binom{m}{\bm} P_{\bm}
\end{eqnarray*}
by the definition (\ref{Pn}) of $P_{\bm}$.  This proves
Proposition~\ref{uniform2}.


Finally, we derive Proposition~\ref{canonical} from
Proposition~\ref{uniform2}. Define $W=T\otimes I + I \otimes T +
V$.  Then the Hamiltonian (\ref{meanFieldHam}) can be written
\[
    H_n \equals \frac{1}{n-1}\sum_{1 \le i < j \le n} W_{ij}\ .
\]
We claim that
\begin{equation}
\label{claim}
    \lim_{n\rightarrow \infty} \frac{n^{-j}}{\tr\Sigma_n} \{ (H_n)^j \Sigma_n\}_{:m}
    \equals
    2^{-j} \big\{ W_{m+1,m+2}W_{m+3,m+4}\cdots W_{m+2j-1,m+2j} S_{m+2j}\big\}_{:m}
\end{equation}
for each $j,m \in \NNN$.  This is so because $(H_n)^j$ contains
$\binom{n}{2}^j$ terms of the form $(n-1)^{-j}W_{a_1 b_1} W_{a_2
b_2}\cdots W_{a_j b_j}$, and, when $n$ is large, the majority of
these terms are such that the indices $a_1,b_1,\ldots,a_j,b_j$ are
all distinct and greater than $m$. The sum of the remaining terms
in $(H_n)^j$ is $o(n^j)$ and does not contribute to the limit
(\ref{claim}).   By the symmetry of $\Sigma_n$,
\begin{eqnarray*}
     \big\{ W_{a_1b_1} W_{a_2b_2}\cdots W_{a_jb_j} \Sigma_n \big\}_{:m}
    & = &
    \big\{ W_{m+1,m+2}\cdots W_{m+2j-1,m+2j} \Sigma_n \big\}_{:m}  \\
    & = &
    \big\{ W_{m+1,m+2}\cdots W_{m+2j-1,m+2j} \Sigma_{n:m+2j} \big\}_{:m}
\end{eqnarray*}
if $a_1,b_1,\ldots,a_j,b_j$ are all distinct and greater than $m$.
There are asymptotically $n^{2j}/2$ such terms, so (\ref{claim})
follows from Proposition~\ref{uniform1}.

Now, to prove Proposition~\ref{canonical}, expand
\[
    \frac{1}{\tr\Sigma_n} e^{- \beta n^{-1} H_n}\Sigma_n  \equals
    \sum_{j=0}^{\infty} \frac{1}{j!} (-\beta)^j n^{-j} (H_n)^j \frac{1}{\tr\Sigma_n}\Sigma_n
\]
and take the $m^{th}$ partial trace:
\begin{equation}
\label{series1}
    \frac{1}{\tr\Sigma_n}\big\{ e^{- \beta n^{-1}  H_n}\Sigma_n \big\}_{:m} \equals\sum_{j=0}^{\infty} \frac{1}{j!} (-\beta)^j n^{-j} \left\{ (H_n)^j \frac{1}{\tr\Sigma_n} \Sigma_n \right\}_{:m}\ .
\end{equation}
The $j^{th}$ term of the series in (\ref{series1}) converges to
\[
    (-1)^j\frac{1}{j!} \Big(\frac{\beta}{2}\Big)^j\big\{ W_{m+1,m+2}W_{m+3,m+4}\cdots W_{m+2j-1,m+2j} S_{m+2j}\big\}_{:m}
\]
as $n \longrightarrow \infty$ by (\ref{claim}) and is bounded by
$\frac{1}{j!} \beta^j\|W\|^j$ uniformly in $n$.  Since the series
in (\ref{series1}) are  majorized by the convergent series $\sum_j
\frac{1}{j!} \beta^j\|W\|^j$ and converge term-by-term as $n
\longrightarrow \infty$, it follows that
\begin{equation}
\label{series2}
    \lim_{n \rightarrow \infty}
    \frac{1}{\tr\Sigma_n} \big\{ e^{-\beta n^{-1}   H_n}\Sigma_n \big\}_{:m} \equals
    \sum_{j=0}^{\infty} (-1)^j\frac{1}{j!} \Big(\frac{\beta}{2}\Big)^j \big\{ W_{m+1,m+2}\cdots W_{m+2j-1,m+2j} S_{m+2j}\big\}_{:m}\ .
\end{equation}
Substituting the integral representations
(\ref{integralRepresentationUniform}) for $S_{m+2j}$ into
(\ref{series2}) yields
\begin{eqnarray*}
   &&
    \lim_{n \rightarrow \infty}
    \frac{1}{\tr\Sigma_n} \big\{ e^{-\beta n^{-1} H_n}\Sigma_n \big\}_{:m} \\
    & = &
    \sum_{j=0}^{\infty} \frac{1}{j!} \Big(\frac{-\beta}{2}\Big)^j
    \int_{\Delta_d} \int_{[0,2\pi)^{d+1}}
       \big[\tr\big(W{P_{\bv(p,\theta)} }^{\otimes 2}\big)\big]^j
              { P_{\bv(p,\theta)} }^{\otimes m}
      \sigma(d\theta) \lambda_d(dp) \\
      & = &
         \int_{\Delta_d} \int_{[0,2\pi)^{d+1}}
         e^{ -\beta \tfrac12 \tr(W P_{\bv(p,\theta)} \otimes P_{\bv(p,\theta)} ) }
          { P_{\bv(p,\theta)} }^{\otimes m}
      \sigma(d\theta) \lambda_d(dp)\ .
\end{eqnarray*}
 Proposition~\ref{canonical} follows from the preceding equation and the definition
(\ref{Gibbs}) of $\Gamma_n(\beta)$.

\section{Acknowledgments}
I would like to thank Lucien Le Cam for listening very patiently
to some of this story and for his kind encouragement. This work
was supported by the Austrian START project ``Nonlinear
Schr\"odinger and quantum Boltzmann equations'' of Norbert J.
Mauser (contract Y-137-Tec).

\end{document}